\documentclass[twocolumn,aps,floatfix]{revtex4}
\usepackage{amssymb}
\usepackage{graphicx}

\begin{document}

\title{Free expansion of a Bose-Einstein condensate at the presence of a thermal cloud}

\author{Krzysztof Gawryluk,$^1$ Miros{\l}aw Brewczyk,$^1$ Mariusz Gajda,$^{2,4}$ and 
        Kazimierz Rz\c a\.zewski$\,^{3,4}$}                          

\affiliation{\mbox{$^1$ Wydzia{\l} Fizyki, Uniwersytet w Bia{\l}ymstoku, 
                        ulica Lipowa 41, 15-424 Bia{\l}ystok, Poland}  \\
\mbox{$^2$ Instytut Fizyki PAN, Aleja Lotnik\'ow 32/46, 02-668 Warsaw, Poland} \\
\mbox{$^3$ Centrum Fizyki Teoretycznej PAN, Aleja Lotnik\'ow 32/46, 02-668 Warsaw, 
           Poland}  \\
\mbox{$^4$ WMP-SN\'S, UKSW, Aleja Lotnik\'ow 32/46, 02-668 Warsaw, Poland}  }             

\date{\today}

\begin{abstract}
We investigate numerically the free-fall expansion of a $^{87}$Rb atoms condensate at 
nonzero temperatures. The classical field approximation is used to separate the condensate 
and the thermal cloud during the expansion. We calculate the radial and axial widths of
the expanding condensate and find clear evidence that the thermal component changes the 
dynamics of the condensate. Our results are confronted against the experimental
data.

\end{abstract}

\maketitle

Since the first experimental realization of a Bose-Einstein condensation in dilute
atomic gases \cite{BEC} the measurement techniques based on time-of-flight expansion 
became a powerful method to study ultracold atomic systems. In fact, this kind of 
measurement was used to prove the existence of a condensate. Starting from an axially 
(cigar- or disc-shaped) symmetric atomic cloud it happened after its release from a trap
that the ratio of the axial and radial both condensate and thermal cloud widths 
systematically change during the expansion. Eventually the anisotropy inversion 
for a condensate was observed which was a crucial distinction from the behavior of a thermal 
cloud. The thermal part, in agreement with the classical Maxwell distribution of velocities 
eventually takes a spherical shape. For a small condensate (like the very first rubidium 
condensate consisting of some $2000$ atoms only) the anisotropy inversion is just a direct 
manifestation of the Heisenberg uncertainty principle - more spatial squeeze - higher momenta. 
For larger samples similar inversion is a result of the interaction energy stored anisotropically 
in the trapped condensate. It is worth adding that the free expansion technique was also used 
for degenerate fermionic gases, for instance to probe the superfluidity of strongly interacting 
atomic Fermi mixtures \cite{mixtures} or to measure the p-wave Feshbach resonances for fermionic 
atoms \cite{p-wave}.

The main purpose of this work is to investigate an influence of a thermal cloud on the 
dynamics of an expanding condensate. If such an impact exists another question is whether 
it is restricted only to times just after the release or is it continued over the whole 
expansion time. Finally, it would be interesting to know how the influence during the
expansion compares to the influence while the system is confined.

To investigate the mutual interaction between the condensed and thermal components during 
the expansion we employ the classical field approximation in a version described in Ref. 
\cite{przeglad}. So, we start with $N$-particle Hamiltonian written in terms of the field 
operator ${\hat {\Psi }({\rm {\bf r}}},t)$ satisfying the bosonic commutation relations.
Assuming the usual contact interaction potential for colliding atoms the Hamiltonian
takes the form:
\begin{eqnarray}
H&=&\int d^3r \, \hat {\Psi }^+({\rm {\bf r}},t)
\left[ -\frac{\hbar^2}{2m} \nabla^2 + V_{tr}({\rm {\bf r}},t)    \right]
\hat {\Psi }({\rm {\bf r}},t)   \nonumber  \\
&+&\frac{g}{2} \int d^3r \, \hat {\Psi }^+({\rm {\bf r}},t) \hat{\Psi }^+({\rm {\bf r}},t) 
\hat {\Psi }({\rm {\bf r}},t)\hat {\Psi }({\rm {\bf r}},t)   \,,
\label{Hamil}
\end{eqnarray}
where the interaction strength  $g=4\pi \hbar^2 a/m$ and $a$ is the s-wave scattering length. 
The trapping potential $V_{tr}({\rm {\bf r}},t)$ is time-dependent and is switched off 
instantaneously to trigger the expansion. The main equation of the classical field approximation 
reads:
\begin{eqnarray}
&&i\hbar \frac{\partial}{\partial t} {\Psi }({\rm {\bf r}},t) =
\left[ -\frac{\hbar^2}{2m} \nabla^2 + V_{tr}({\rm {\bf r}},t)    \right]
{\Psi }({\rm {\bf r}},t)   \nonumber  \\
&&+ g\, {\Psi }^*({\rm {\bf r}},t) {\Psi }({\rm {\bf r}},t)
{\Psi }({\rm {\bf r}},t)  
\label{CFequation}
\end{eqnarray}
and is just the Heisenberg equation of motion for the field operator stripped of its
operator character. The complex wave function ${\Psi }({\rm {\bf r}},t)$ which we call the
classical field describes both condensed and noncondensed atoms. The use of the classical field
instead of the field operator is justified when only macroscopically occupied modes are
taken into consideration. This reasoning remains in analogy with the treatment of an intense
light beam which although consisting of single photons maybe described by the electric and 
magnetic fields.

An important question how to get out of the classical field the information on a condensate
and a thermal cloud is resolved by using of Penrose and Onsager definition of a Bose-Einstein
condensation \cite{POdef} and by taking into consideration a measurement process. Since any
detector has a limited spatial and temporal resolutions a complicated (both in space and time)
behavior of the high energy classical field is smoothed out during the measurement. Therefore,
the quantity which is physically important is a time and/or space averaged one-particle
density matrix. According to the Penrose and Onsager definition the condensate wave function
is an eigenvector corresponding to the dominant eigenvalue of a one-particle coarse-grained
(i.e., averaged over time and/or space) density matrix. We closely follow the experiment
and realize the averaging as a column integration along one of the radial directions. The 
physically important (averaged along $y$ direction) one-particle density matrix is given by
\begin{equation}
\bar{\rho}(x,z,x',z';t) = \int dy \, \Psi(x,y,z,t) \, \Psi^*(x',y,z',t) 
\label{rhoave}
\end{equation}
and the splitting procedure requires the diagonalization of (\ref{rhoave}). This kind
of averaging was already used to investigate a decay of multiply charged vortices
\cite{vortices}. The splitting procedure is then summarized as:
\begin{eqnarray}
&&\bar{\rho}=\sum_k N_k \, \varphi_k(x,z,t) \, \varphi^*_k(x^{\,\prime},z^{\,\prime},t)
\\
&&\psi_0(x,z,t) =  \sqrt{N_0}\, \varphi_0(x,z,t)
\\
&&\rho_T(x,z,t) = \bar{\rho}(x,z,x,z;t) - |\psi_0(x,z,t)|^2  \,.
\end{eqnarray}
Here, $\varphi_k$ are the macroscopically occupied modes, $N_0$ is the dominant eigenvalue,
$\psi_0$ is the condensate wave function, and $\rho_T$ is the density of thermal cloud.

Having introduced the classical field approximation we now describe our numerical procedure.
First, we find the classical field corresponding to the $^{87}$Rb Bose gas (with a scattering
length $a=5.82\,$nm) at equilibrium confined in a harmonic trap with frequencies 
$\omega_{\bot} \equiv \omega_{x,y} = 2\pi \times 137.4\,$Hz and 
$\omega_z = 2\pi \times 12.6\,$Hz. Details on how to obtain an equilibrium state for a given
number of atoms and at particular temperature are explained elsewhere \cite{oscylacje}.
Since we intend to investigate the influence of thermal atoms on the condensate expansion we 
prepare various equilibrium states but we keep the same number of condensed atoms ($30000$
or $90000$) in all of these states. 
Next, we suddenly turn off the trapping potential and let the atomic cloud to
expand. Technically speaking, we solve the Eq. (\ref{CFequation}) on a larger grid (but having 
the same spatial step) and without any trap. We monitor the momentum distribution (i.e.,
the Fourier transform of the classical field) during the expansion and find that it changes
for the first few milliseconds only. In other words, a few milliseconds is required to convert fully 
the interaction energy into the kinetic energy. Afterwords, the classical field evolves freely
and can be found with the help of the propagator of the free Schr\"odinger equation.
Finally, at a desired time the splitting of the classical field into the condensed and 
noncondensed components is performed.

\begin{figure}[htb]
\resizebox{2.5in}{3.25in} {\includegraphics{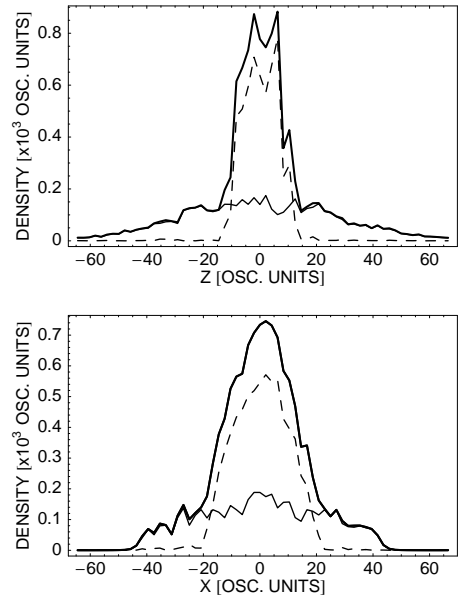}}
\caption{Axial (upper frame) and radial (lower frame) cuts of the total (thick solid line), 
condensate (dashed line), and thermal (thin solid line) densities as obtained by splitting the 
free expanding classical field at $22\,$ms as described in the text. The condensate fraction 
is equal to $0.3$ and the number of condensed atoms $N_0 = 90000$. The oscillatory unit of 
length is defined based on the axial trap frequency: 
$\sqrt{\frac{\hbar}{m \omega_z}}$ and equals $3.0\, \mu m$. }
\label{densities}
\end{figure}

In Fig. \ref{densities} we plot the radial and axial densities of an expanding atomic cloud 
at $22\,$ms. Initially the condensate is a cigar-shaped like a trap. Its aspect
ratio in the Thomas-Fermi approximation is given by 
$R_{\bot}/R_z = \omega_z /\omega_{\bot}$ \cite{Pethick} which equals approximately
$1/10$. After $22\,$ms, as can be seen in Fig. \ref{densities}, the initial
anisotropy is, actually, inverted. The radial size gets larger than the axial one. This, 
of course, is not true for the thermal cloud in which case the final density becomes 
spherical.

\begin{figure}[htb]
\resizebox{3.5in}{2.3in} {\includegraphics{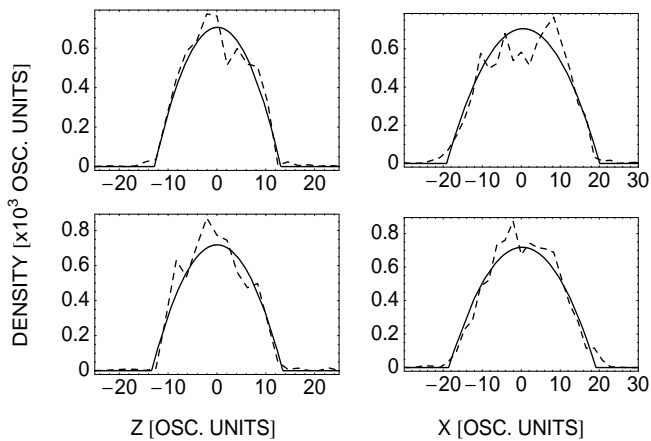}}
\caption{Condensate axial (left frame, dashed line) and radial (right frame, dashed line) 
densities for two single realizations (upper and lower frames, respectively) at $22\,$ms. 
Condensate density is extracted from the classical field by the splitting procedure and is 
fitted to two-dimensional inverse parabola. Black lines show axial (left frame) and radial 
(right frame) cuts of such a fit. The parameters are the same as in Fig. \ref{densities}.}
\label{realization}
\end{figure}

Fig. \ref{realization} reveals some technical details related
to the read-out procedure. After the splitting of the classical field is concluded
and the two-dimensional condensate density is known, this density is fitted by a 
two-dimensional inverted parabola (since according to the large number of atoms in the
condensate the Thomas-Fermi approximation is valid). The fit is performed based on
the least squares method. In fact, fits depend on the realization and, as will be shown 
later, the aspect ratio is a quantity which is most sensitive to the realization.

\begin{figure}[htb]
\resizebox{3.2in}{2.3in} {\includegraphics{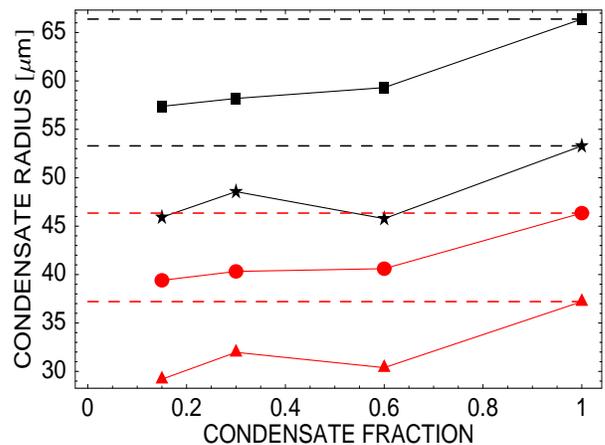}}
\caption{(Color online) Radial (squares and stars) and axial (circles and triangles) 
condensate radii as a function of the condensate fraction after $22\,$ms of ballistic 
expansion. Two sets of points marked by squares and circles correspond to the systems 
with the number of condensed atoms $N_0=90000$ whereas two other sets (marked by stars 
and triangles) represent the systems with $N_0=30000$ condensed atoms. Solid lines are 
shown to guide the eye. Horizontal dashed lines are the Thomas-Fermi values for the radial
lengths (two upper lines, the upper line for $N_0=90000$ and the lower line for
$N_0=30000$) and the axial lengths (two lower lines, the upper line for $N_0=90000$ and 
the lower line for $N_0=30000$).
Note that both radial and axial widths get shorter in comparison with the size of a pure 
condensate of the same number of atoms. }
\label{radii}
\end{figure}

The main result of this work is presented in Fig. \ref{radii}. It shows the radial
and axial condensate widths after $22\,$ms of free expansion as a function of a
condensate fraction. There are two sets of data included in this figure. The first one 
(squares and circles) depicts the behavior of the system with the same number of condensed 
atoms ($N_0=90000$) independently of the condensate fraction (i.e., temperature). The
second set (starts and triangles) represents data for the systems with $N_0=30000$.
Horizontal lines are the radial and axial sizes of $\,9\times 10^{4}$ and $3\times 10^{4}$
atoms condensate calculated within the expansion model formulated by Castin and Dum 
\cite{TF_CD}. This model describes the free expansion of a pure condensate within the
Thomas-Fermi limit. Our data suggest that thermal atoms somehow temper the expansion of 
the condensate.

There are three experimental papers discussing the temperature effects having influence 
on the ballistic expansion of a condensate \cite{Aspect,Zawada,Bagnato}. All of these 
papers claim that the behavior of the condensed cloud measured in terms of its size 
during the expansion depends on the temperature of the system before the expansion. 
So, the quantitative comparison between the numerical calculations and the experiment 
is possible.

\begin{figure}[htb]
\resizebox{3.5in}{2.6in} {\includegraphics{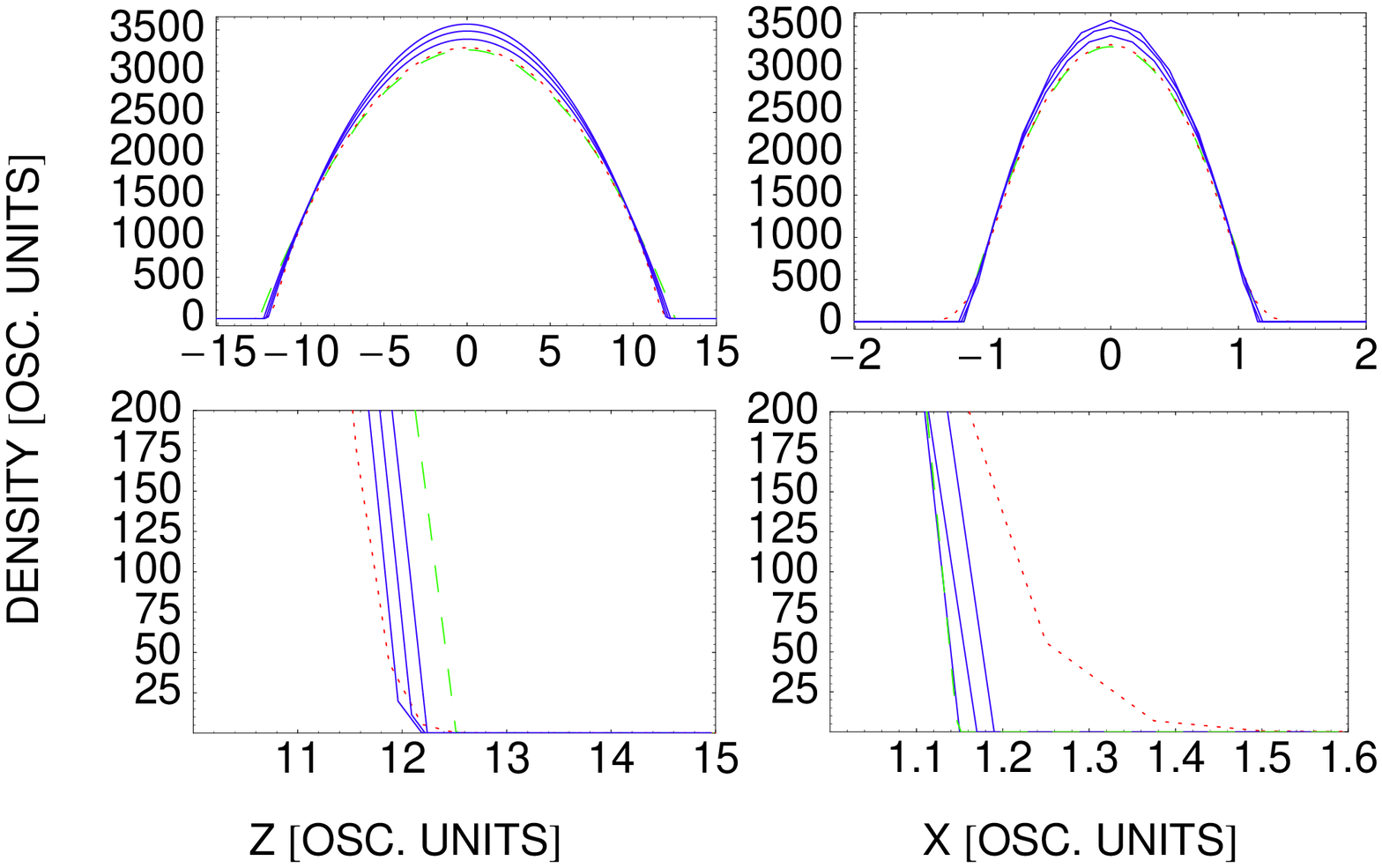}}
\caption{(Color online) Axial (left frames) and radial (right frames) condensate density 
cuts for a system with $N_0=90000$ condensed atoms calculated within the self-consistent 
Hartree-Fock model (solid lines corresponding to the condensate fractions $0.15$, $0.30$, 
and $0.60$), by the Gross-Pitaevskii equation (dotted line), and from the Thomas-Fermi 
formula (dashed line). Lower panel shows in detail the region the density drops to zero. }
\label{HFden}
\end{figure}

Since our numerical parameters were taken in a way to match the parameters of experiment 
of Ref. \cite{Zawada} we start with this paper. In that experiment the authors make effort 
to keep constant the number of condensed atoms while expanding the atomic samples at various 
temperatures. They found the increase of radial and axial condensates lengths when the 
temperature gets higher (see Fig. 4 in \cite{Zawada}). The measured widths are larger than 
the corresponding Castin-Dum values. So, the authors claim that their experiment is performed 
in the non-Thomas-Fermi regime. An increase of radial and axial widths with temperature 
is explained by an assumption that at equilibrium in a trap the thermal atoms exert
a force on condensed atoms towards the center of a trap thus compressing the condensate
cloud. This compression results in a faster expansion in all directions after the trap
is released. However, our calculations within the classical field approximation show that, 
actually, no compression of a condensate cloud occurs in a trap. We stress that this 
statement is true also within the self-consistent Hartree-Fock model \cite{Pethick}. 
In Fig. \ref{HFden} we plot axial and radial condensate densities obtained by solving 
selfconsistently the equations of the Hartree-Fock model for a particular number of 
condensed atoms ($N_0=90000$) but at various temperatures (solid lines which correspond 
to the condensate fractions $0.15$, $0.30$, and $0.60$). We also added to the figure the 
densities of a pure condensate consisting of $90000$ atoms calculated by solving the 
Gross-Pitaevskii equation in imaginary time (dotted line) as well as by the Thomas-Fermi 
formula (dashed line). Clearly, Fig. \ref{HFden} shows no compression due to the presence 
of a thermal cloud.

\begin{figure}[htb]
\resizebox{3.in}{3.5in} {\includegraphics{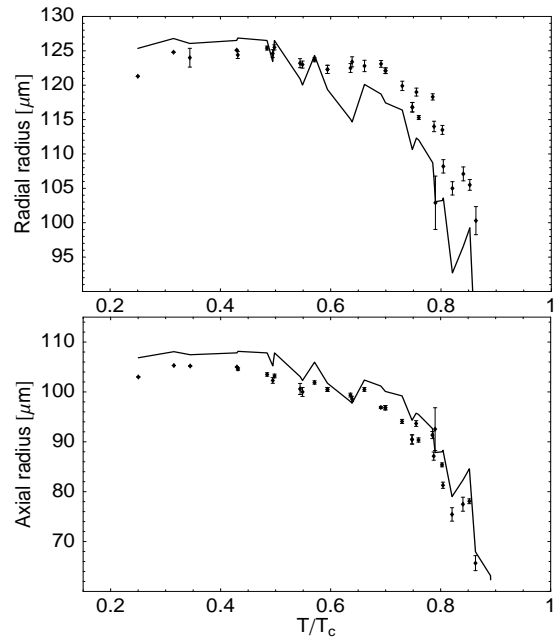}}
\caption{Radial (upper panel) and axial (lower panel) widths as a function of reduced
temperature. Experimental data (points with error bars) correspond to the experiment 
of Ref. \cite{Aspect} whereas numerical results (solid lines) are calculated based on 
the Castin-Dum approximation. }
\label{GA}
\end{figure}

A qualitative difference between numerical results and the experimental data of
Ref. \cite{Zawada} motivated us to make a comparison with other experimental works.
For example, in Ref. \cite{Aspect} a deviation from ballistic expansion is also 
reported. In Fig. 5a of that paper the authors plot the aspect ratio of the condensed 
component after $22.3\,$ms of free expansion as a function of reduced temperature. 
It is clear from this figure that for higher temperatures the results are different
than the Castin-Dum limit \cite{TF_CD} (the aspect ratio shows deviation from the
Castin-Dum values also in \cite{Zawada}). Therefore, we look separately at the axial 
and radial sizes of the expanding condensate in the case of Orsay experiment. The results 
are presented in Fig. \ref{GA}. Here, the experimental data are compared with the 
Castin-Dum values. This figure clearly shows that the experimental data stay close to
the Castin-Dum values. In axial direction the thermal cloud seems to temper the expansion
of a condensate whereas the interplay between the condensate and the thermal component
in radial direction gets more complicated. Differences are on the level of a few percent 
similarly to what we obtain from our numerics (although for a different trap geometry) 
and in opposite to what is reported in paper \cite{Zawada}.

\begin{figure}[htb]
\resizebox{3.2in}{2.3in} {\includegraphics{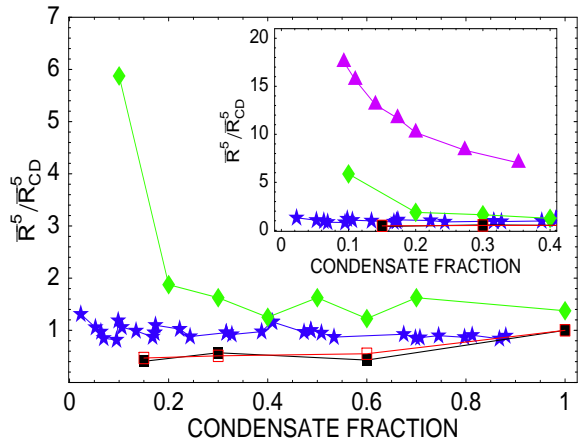}}
\caption{(Color online) Ratio $\bar{R}^5/N_0$ (normalized to the value obtained from the 
Castin-Dum approximation) as a function of the condensate fraction. The mean size of the 
condensate is defined as $\bar{R}=(R_{\bot}^2 R_z)^{1/3}$. Four sets of data are presented 
in the figure according to the numerical calculations (solid and open squares), the results
of Ref. \cite{Aspect} (stars), Ref. \cite{Bagnato} (diamonds), and Ref. \cite{Zawada} 
(inset, triangles). }
\label{R5N0}
\end{figure}

In other experimental work, Ref. \cite{Bagnato}, the authors find finite temperature
correction to the Thomas-Fermi approximation as a function of the condensate fraction 
by measuring the ratio $\bar{R}^5/N_0$, where $\bar{R}$ is the condensate radius
defined as $\bar{R} = (R_{\bot}^2 R_z)^{1/3}$ and $N_0$ is the number of condensed atoms.
There is an agreement with Castin-Dum predictions \cite{TF_CD} for low temperatures,
however, when the temperature gets higher the ratio $\bar{R}^5/N_0$ departs from the
Castin-Dum value getting larger (see Fig. 4 of \cite{Bagnato}). The authors explain
this behavior by using a combination of a modified Hartree-Fock model to describe the 
condensed and thermal fractions in a trap and an expansion model formulated by 
Castin and Dum \cite{TF_CD}. They conclude that the influence of the thermal cloud on the 
condensate during the expansion is negligible which seems to be in opposition to what is
claimed in Refs. \cite{Aspect} and \cite{Zawada}. It contradicts also our findings.
Therefore, we decided to compare all experiments and our numerical results on a graph
where we plot $\bar{R}^5/N_0$ (actually, normalized to the value given by the Castin-Dum
approach to make the comparison feasible) as a function of the condensate fraction.
In Castin-Dum formulation one has

\begin{eqnarray}
&& \bar{R}^5(t)/N_0 = 15\, a\, \bar{a}^4  (\lambda_{\bot}^2(t)\, \lambda_z(t))^{5/3}
\end{eqnarray}
where
\begin{eqnarray}
&& \lambda_{\bot}(t) = \sqrt{1+t^2}   \nonumber \\
&& \lambda_z(t) = 1+\beta^2 (t\, \arctan{t}-\ln{\sqrt{1+t^2}})  \, ,
\end{eqnarray}
$\beta=\omega_z/\omega_{\bot}$ and time $t$ is expressed in units of $1/\omega_{\bot}$
whereas $\bar{a}$ is an oscillatory unit length calculated based on the geometric
mean of all angular frequencies. Fig. \ref{R5N0} shows that our results (solid
squares for $N_0=90000$ and open squares for $N_0=30000$) stay in a quite good agreement 
with Orsay experimental data.

Certainly, further experimental and theoretical effort is required to gain more
insight to what, indeed, is happening during the expansion of condensate and thermal
cloud.

\begin{figure}[htb]
\resizebox{3.2in}{2.1in} {\includegraphics{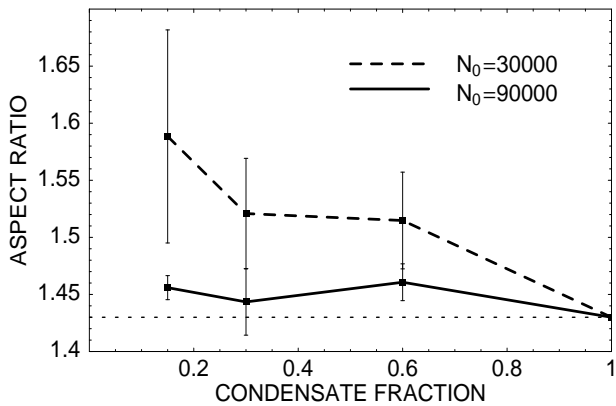}}
\caption{Aspect ratio ($R_{\bot}/R_z$) at $22\,$ms defined as the ratio of radial and 
axial condensate widths for various condensate fractions. Two sets of data are presented 
corresponding to the number of condensed atoms $N_0=90000$ (solid line) and $N_0=30000$ 
(dashed line). Error bars depict the dependence on the realization. Horizontal dashed 
(thin) line is the aspect ratio obtained within the Thomas-Fermi approximation. }
\label{ar}
\end{figure}

Finally, in Fig. \ref{ar} we plot the aspect ratio for a condensate for various
condensate fractions. The aspect ratio seems to be a quantity which is most sensitive
to the interplay between the thermal cloud and the condensate. For low temperatures 
the aspect ratio approaches the Castin-Dum value whereas for higher temperatures it is 
getting larger. Moreover, we find that the aspect ratio is very sensitive to the initial 
conditions fulfilled by the classical field. Different realizations lead to different 
aspect ratios what is marked by error bars in Fig. \ref{ar}.

In conclusion, we have studied the expansion of the Bose-Einstein condensate at the presence
of thermal atoms. Using the classical field approximation we have shown that thermal atoms 
change the dynamics of a condensate in a way 
that both radial and axial condensate widths get smaller in comparison with the case when
there is no thermal cloud. It results in a change of condensate aspect ratio which
becomes bigger for smaller condensate fraction (i.e., larger thermal cloud). While all papers 
agree that the thermal cloud does play a role in the expansion of the condensate, the details 
remain unclear. The three experimental papers are not mutually in agreement and also our results 
do not coincide with some measurements. Clearly more work is needed to clarify this somewhat 
confusing situation.

\acknowledgments
We are grateful to all three experimental groups (Orsay, Toru\'n, and S\~ao Carlos) for helpful 
discussions and for providing us with their raw data.
The authors acknowledge support by Polish Government research funds for 2009-2011.
Some of the results have been obtained using computers at the Interdisciplinary
Centre for Mathematical and Computational Modeling of Warsaw University.

\end{document}